\documentclass[allclo]{FBSart}
\usepackage{amsfonts}
\usepackage{amssymb}
\usepackage{epsfig}
\title{A dispersive approach to pion photo- and electroproduction}
\author{D. Drechsel$^{1)}$
\thanks{\textit{E-mail address:} drechsel@kph.uni-mainz.de}, B. Pasquini$^{2)}$, and L. Tiator$^{1)}$}
\institute{$^{1)}$ Institut f{\"u}r Kernphysik, Johannes
Gutenberg-University{\"a}t,
55099 Mainz, Germany\\
$^{2)}$ Dipartimento di Fisica Nucleare e Teorica, University\`{a}
degli Studi di Pavia and\\ \quad INFN, Sezione di Pavia, Pavia,
Italy}

\runningauthor{D. Drechsel, B. Pasquini, and L. Tiator} \runningtitle{pion
photo- and electroproduction}

\begin{document}

\maketitle
\begin{abstract}
{The relativistic amplitudes of pion photo- and electroproduction are
calculated by dispersion relations at constant $t$. Several sum rules and
low-energy theorems for the threshold amplitudes are investigated within this
technique. The continuation of the amplitudes to sub-threshold kinematics is
shown to provide a unique framework to derive the low-energy constants of
chiral perturbation theory by global properties of the excitation spectrum.}
\end{abstract}
\section{Introduction}
In two recent publications we have studied pion photoproduction on the nucleon
in the framework of fixed-$t$ dispersion relations~\cite{Pas05,Pas06}. In
particular, we have concentrated on the threshold region in which the results
can be compared to both precision data and predictions of baryon chiral
perturbation theory (ChPT). The dispersion relations (DRs) are based on a set
of 4 photoproduction amplitudes $A_i(\nu ,t)$ depending on energy and momentum
transfer described by the Lorentz invariant variables $\nu$  and $t$,
respectively. These relations are Lorentz and gauge invariant by construction,
and unitarity is implemented by constructing the real parts of the amplitudes
from the imaginary (absorptive) parts via the dispersion integrals. The
dispersive amplitudes for sub-threshold kinematics are regular functions in a
region of small $\nu$ and $t$ values, and therefore they can be expanded in a
power series about the origin of the Mandelstam plane ($\nu =0 ,\, t=0$).
Comparing this series with the tree and loop contributions of relativistic
baryon ChPT~\cite{Ber92a,Ber94,Ber05} one can read off the required low-energy
constants (LECs) of that field theory, which up to now have been fixed by
resonance saturation models or fits to the threshold data. In our present work
we use MAID05~\cite{MAID} as input for the absorptive parts of the amplitudes,
which are obtained over the full resonance region up to c.m. energies of
W=2.2~GeV by a global fit to the pion photoproduction data. With few exceptions
the results compare favorably with the experimental data and the predictions of
ChPT in the threshold region.

Another interesting aspect is the comparison with sum rules and low-energy
theorems (LETs) of the 1950's and 1960's. These relations were based on current
algebra and the PCAC hypothesis (partial conservation of the axial current).
They become exact in the chiral limit of QCD, and thus all variables and
observables have to be understood in the fictitious limit of vanishing (light)
quark masses and hence soft pions with mass $M_{\pi} \rightarrow 0$. This leads
to an expansion of the amplitudes in the mass ratio $\mu =M_{\pi}/M_N$. The
first such theorem was established by Kroll and Ruderman~\cite{Kro54} for
charged pion photoproduction, with the result that the S-wave amplitude for
this process was finite in the limit $\mu \rightarrow 0$, i.e., ${\mathcal
{O}}(1)$. Somewhat later, several authors derived a LET for neutral pion
photoproduction~\cite{Vai70,deB70}, which led to the assumption that this
reaction was ${\mathcal {O}}(\mu)$ for the proton and ${\mathcal {O}}(\mu^2)$
for the neutron. However, extensive investigations in ChPT have shown that the
finite pion mass leads to substantial corrections at physical
threshold~\cite{Ber91,Ber92}. We have studied the role of such corrections in
the context of two sum rules of Fubini, Furlan, and Rossetti
(FFR)~\cite{Fub65}. The first of these sum rules relates the nucleon's
anomalous magnetic moment $\kappa_N$ to a dispersion integral over the first
pion photoproduction amplitude $A_1 (\nu , t)$ for neutral pions, or, if
extended to electroproduction, the Pauli form factor to the corresponding
electroproduction amplitude. The second FFR sum rule predicts that the
difference of the axial ($G_A^V$) and the Dirac vector ($F_1^V$) form factors
is related to the longitudinal electroproduction amplitude $A_6$ for charged
pions. The radii of these form factors are of similar size, $\langle r^2
\rangle_1^V - \langle r^2 \rangle_A^V=(0.14 \pm 0.03)$ fm$^2\,$ and the early
estimates~\cite{Ria66} simply led to the result $G_A^V(Q^2)-F_1^V(Q^2)=0$. The
first and to our knowledge only dispersive calculation was performed by Adler
and Gilman already in 1966~\cite{Adl66}. Their result was $\langle r^2
\rangle_1^V - \langle r^2 \rangle_A^V=0.152 $ fm$^2$, in fantastic agreement
with our present knowledge of this observable. Unfortunately, this result
involves a dispersion integral with formidable cancelations (I) among
contributions of positive sign in the region up to the $\Delta (1232)$
resonance and of negative sign in the second resonance region and (II) between
the electric transverse and longitudinal contributions of the same
multipolarity.

We proceed by reviewing the status of the LET for neutral pion photoproduction
in Sec.~2. The formalism for pion electroproduction is briefly described in
Sec.~3. In the following Sec.~4 we discuss the FFR sum rule for the anomalous
magnetic moment and compare the results of our dispersive calculation with the
predictions of ChPT. We present some preliminary results on the two FFR sum
rules for electroproduction in Sec.~5, and close by a short summary in Sec.~6.
\section{Revised LET for neutral pion photoproduction}
The ``LET'' of Refs.~\cite{Vai70,deB70} for the reaction $\gamma+p\rightarrow
\pi^0+p$ was based on current algebra and PCAC. According to the theorem, the
leading terms of the threshold multipole were to be given by the Born diagrams,
evaluated with the pseudovector pion-nucleon interaction. However, this
prediction had to be revised in the light of surprising experimental evidence.
The reason for the discrepancy between the theorem and the data was first
explained in the framework of ChPT by pion-loop corrections. An expansion of
the S-wave amplitude in the mass ratio $\mu$ yielded the
result~\cite{Ber91,Ber92}
\begin{equation}
E_{0+} (\pi^0 p) = \frac{eg_{\pi N}}{8\pi m_{\pi}} \left \{ \mu-\mu^2\,
\frac{3+\kappa_p}{2}- \mu^2\, \frac{M^2}{16f^2_{\pi}} +\ ...\right \}\ ,
\label{eq21}
\end{equation}
where $g_{\pi N}$ is the pion-nucleon coupling constant and $f_{\pi} \approx
93$~MeV the pion decay constant. The first and the second terms on the rhs of
Eq.~(\ref{eq21}) are the prediction of Refs.~\cite{Vai70,deB70}, which however
has to be corrected by the third term on the rhs. Although this loop correction
is formally of higher order in $\mu$, its numerical value is of the same size
as the leading term.
\begin{figure}[hbt]
\begin{center}
\epsfig{file=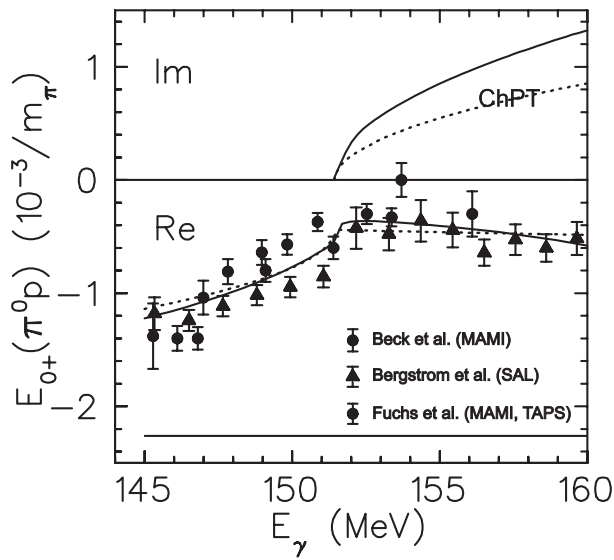,width=8cm}
\end{center}
\caption[]{The real (Re) and imaginary (Im) parts of the S-wave amplitude
$E_{0+}$ for $\pi^0$ photoproduction at threshold energies. The MAMI data
\cite{Bec90,Fuc96} are represented by circles, the SAL data \cite{Ber96} by
triangles. Dashed lines: predictions of ChPT at $\cal{O}$($p^3$)
\cite{BKM96,BKM96a}, solid lines: results from dispersion relations
\cite{Han96}. The solid horizontal line at about -2.2 shows the prediction of
Refs.~\cite{Vai70,deB70}.} \label{fig21}
\end{figure}
The energy dependence of $E_{0+}(\pi^0 p)$ is shown in Fig.~\ref{fig21}. The
discrepancy between the prediction of Refs.~\cite{Vai70,deB70} and the
experimental data obtained at the Mainz Microtron MAMI and at SAL (Saskatoon)
is apparent. Furthermore, the real part of the amplitude shows a characteristic
``Wigner cusp'' at the threshold for charged pion production, which lies about
5~MeV above the $\pi^0$ threshold. This cusp in the real part is related to the
sharp rise of the imaginary part at the second threshold. The physical picture
behind the large loop correction is based on (I) the large production rate of
the charged pions and (II) the charge-exchange scattering between the nucleon
and the slow $\pi^+$ in the intermediate state, which leaves a $\pi^0$ in the
final state. However, the direct experimental determination of the imaginary
part will require double-polarization experiments with linearly polarized
photons and polarized targets. The excellent agreement between ChPT and the
data for $E_{0+}$ is somewhat flawed by the fact that higher order diagrams are
sizeable, that is, the perturbative series converges slowly and low-energy
constants appearing at the higher orders reduce the predictive
power.\\
\begin{figure*}[]
\begin{center} \epsfig{figure=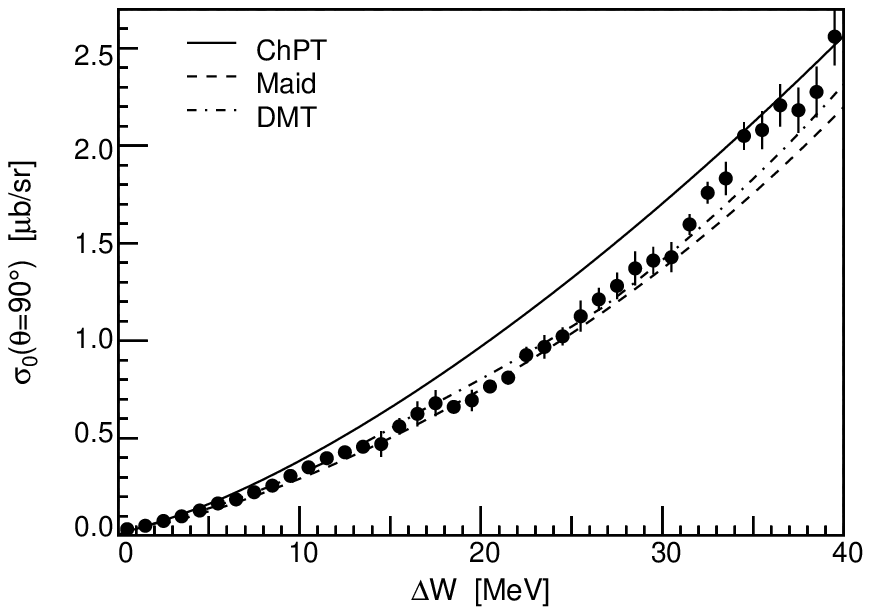,width=6cm,angle=0}
\epsfig{figure=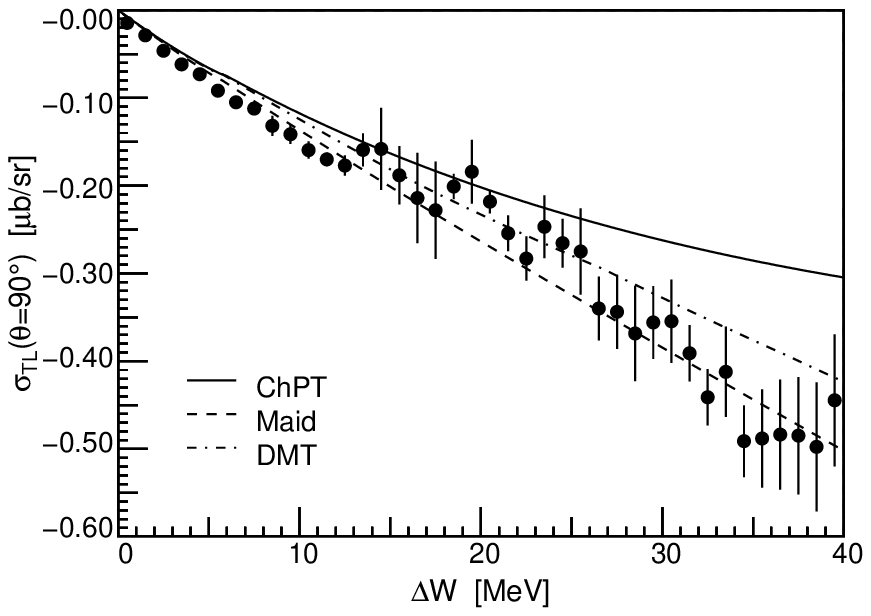,width=6cm,angle=0}\\
\epsfig{figure=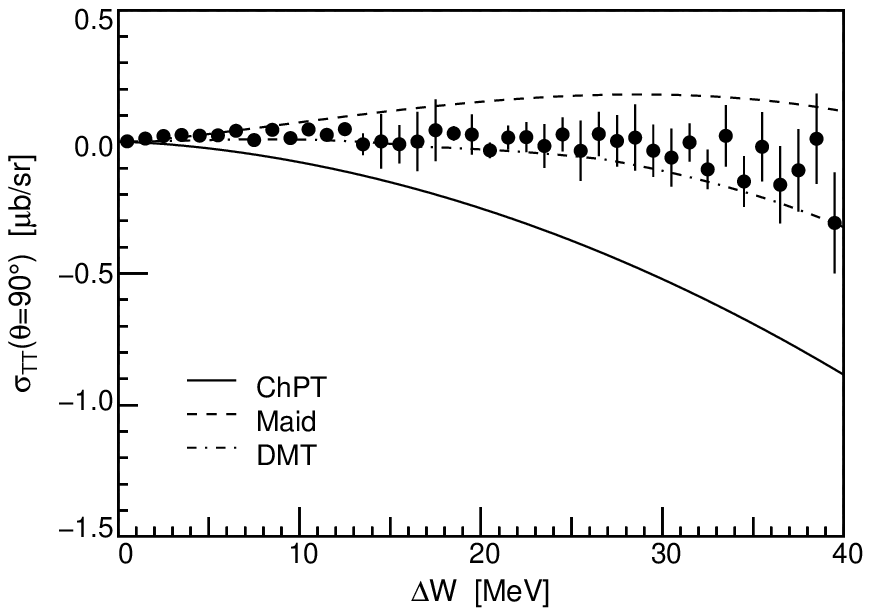,width=6cm,angle=0}
\epsfig{figure=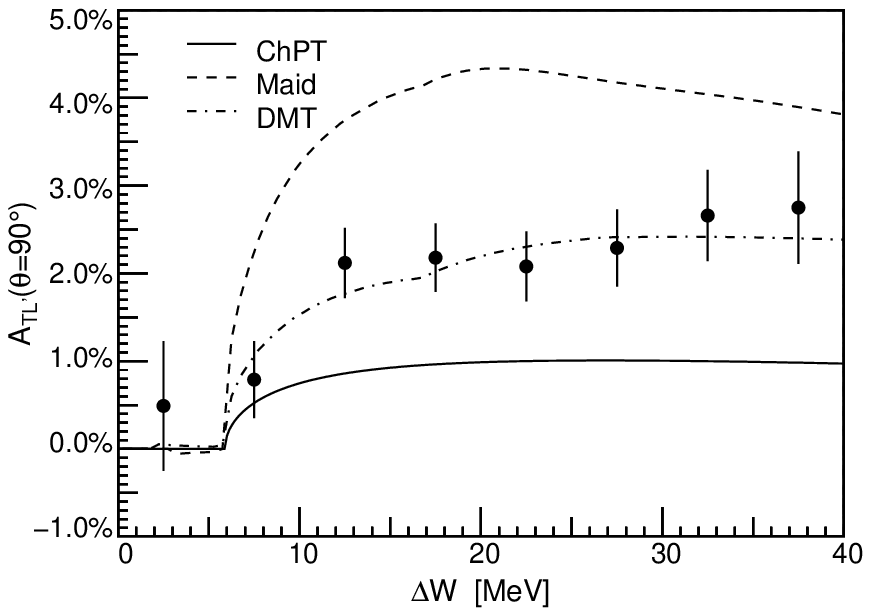,width=6cm,angle=0}
\end{center}
\caption{The separated cross sections $\sigma_0$, $\sigma_{\rm{TL}}$, and
$\sigma_{\rm{TT}}$ as well as the beam helicity asymmetry $A_{\rm{TL}}'$, as
function of $\Delta W=W-W_{\rm {thr}}$ and at the pion c.m. angle
$\theta_{\pi}=90^\circ$. Solid line: HBChPT~\cite{Ber96b}, dashed line:
MAID~\cite{MAID}, and dashed-dotted line: DMT model \cite{Kam01}. The data are
from Weis {\it et al.}~\cite{Wei07}.} \label{fig22}
\end{figure*}
\indent The great success of ChPT for photoproduction at threshold
was a strong motivation to extend the experimental program to
electroproduction. The first of such investigations were performed
at NIKHEF~\cite{vdB95} and MAMI~\cite{Dis98} for $Q^2= 0.10~
$GeV$^2$, and provided another confirmation of ChPT although at the
expense of 2 new low-energy constants, which were fitted to the
data. In order to further check this agreement, data were also taken
at lower momentum transfer. Whereas the former experiments were only
sensitive to the real part of the amplitudes, Weis {\it et
al.}~\cite{Wei07} also determined the fifth structure function
(${\rm{TL}}'$), which contains information on the phase of the
S-wave amplitude. The result is displayed in Fig.~\ref{fig22}. We
observe that only the dynamical Dubna-Mainz-Taipei (DMT) model
\cite{Kam01} is able to fully describe the experiment, in particular
its prediction for the helicity asymmetry is right on top of the
data. Such dynamical models start from a description of the
pion-nucleon scattering phases by a quasi-potential, which serves as
input for an integral equation to account for multiple scattering.
In this sense the model contains the loop corrections to an
arbitrary number of rescattering processes, and is therefore
perfectly unitary, albeit on a phenomenological basis that may
violate gauge invariance to some extent.
\section{Formalism}
\begin{figure}[hbt]
\begin{center}
\epsfig{file=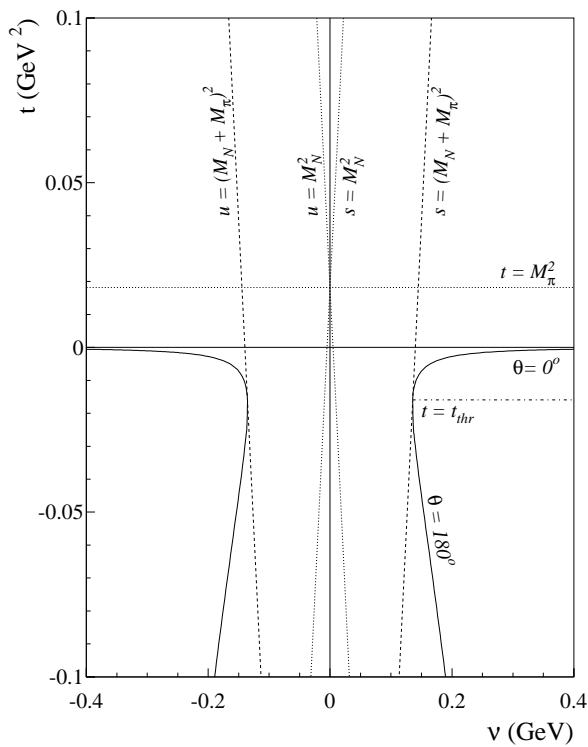,width=8cm} \caption[]{The Mandelstam plane for
pion photoproduction ($Q^2=0$) on the nucleon. The boundaries of the physical
region are $\theta=0$ (forward production) and $\theta=180^{\circ}$ (backward
production). The nucleon and pion pole positions are indicated by the dotted
lines $s=M_N^2$, $u=M_N^2$, and $t=M_{\pi}^2$. The dashed line
$s=(M_N+M_\pi)^2$ indicates the threshold of pion production and therefore also
of the imaginary part of the production amplitude. It is tangent to the
boundary of the physical region in the point $\nu=\nu_{\rm{thr}}$,
$t=t_{\rm{thr}}$. The path $t=t_{\rm{thr}}$ (dashed-dotted line) yields the
only dispersion relation at $t={\rm{const}}$ whose imaginary part is fully
contained in the physical region.}
\end{center}
\label{fig31}
\end{figure}

Let us consider the reaction $\gamma^{\ast}(q) + N (p) \rightarrow \pi(k) +
N(p\,')$, where the variables in brackets denote the four-momenta of the
participating particles. The familiar Mandelstam variables are
$s=(p+k)^2,\,t=(q-k)^2$, and $u=(p-q)^2$, and $\nu=(s-u)/(4M_N)$ is the
crossing-symmetric variable. The latter variable is related to the photon lab
energy $E_\gamma^{\rm{lab}}$ by $\nu=E_\gamma^{\rm{lab}} +
(t-M_{\pi}^2+Q^2)/(4M_N)$. The physical $s$-channel region is shown in
Fig.~\ref{fig31}. Its upper and lower boundaries are given by the scattering
angles $\theta=0$ and $\theta=180^{\circ}$, respectively. The nucleon and pion
poles lie in the unphysical region and are indicated by the dotted lines at
$\nu_s=\nu_B$ ($s$-channel) and $\nu_u=-\nu_B$ ($u$-channel), where
$\nu_B =(t-M_\pi^2+Q^2)/(4M_N)$, and $t=M_{\pi}^2$ ($t$-channel).\\
\indent The nucleonic transition current can be expressed in terms of 6
invariant amplitudes $A_i$, $J^\mu = \sum_{i=1}^6 A_i(\nu,t,Q^2)\, M^\mu_i$,
with the four-vectors $M^\mu_i$ given by the independent axial vectors
constructed from the particle momenta and the Dirac $\gamma$
matrices~\cite{Den61}. In the case of real photons ($Q^2=0$) and with the gauge
condition $\epsilon_\mu k^\mu=0$, the matrices $M^\mu_5$ and $M^\mu_6$ do not
contribute to the interaction Lagrangian, and the remaining 4 matrices reduce
to Eq.~(10) of Ref.~\cite{Pas05}. The invariant amplitudes $A_i$ can be further
decomposed into three isospin channels ($a=1,2,3$),
$A_i^a=A_i^{(-)}i\epsilon^{a3b}\tau^b+A_i^{(0)}\tau^a+A_i^{(+)}\delta_{a3}$,
where $\tau^a$ are the Pauli matrices in isospace, and the physical
photoproduction amplitudes are given by
\begin{eqnarray}
A_i(\gamma p\rightarrow n\pi^+)=+\sqrt{2}(A_i^{(-)}+A_i^{(0)}),&&
A_i(\gamma p\rightarrow p\pi^0)=A_i^{(+)}+A_i^{(0)},\nonumber\\
A_i(\gamma n\rightarrow p\pi^-)=-\sqrt{2}(A_i^{(-)}-A_i^{(0)}),&& A_i(\gamma
n\rightarrow n\pi^0)=A_i^{(+)}-A_i^{(0)}. \label{eq31}
\end{eqnarray}
Under crossing, the amplitudes $A_{1,2,4}^{(+,0)}$ and
$A_{3,5,6}^{(-)}$ are even functions of $\nu$ and satisfy a DR of
the type
\begin{equation}
{\rm Re}\,A^{I}_i(\nu,t,Q^2)= A_{i,{\rm pole}}^I(\nu,t,Q^2) +\frac{2}{\pi}{\cal
P}\int_{\nu_{\rm{thr}}}^{\infty}{\rm d}\nu'\, \frac{\nu'\,{\rm
Im}\,A_i^I(\nu',t,Q^2)}{\nu'^2-\nu^2}\,, \label{eq32}
\end{equation}
whereas the amplitudes $A_{3,5,6}^{(+,0)}$ and $A_{1,2,4}^{(-)}$ are
odd and therefore fulfil the relation
\begin{equation}
{\rm Re}\,A^I_i(\nu,t,Q^2)=A_{i,{\rm pole}}^I(\nu,t,Q^2) +\frac{2\nu}{\pi}{\cal
P}\int_{\nu_{\rm{thr}}}^{\infty}{\rm d}\nu' \, \frac{{\rm
Im}\,A_i^I(\nu',t,Q^2)}{\nu'^2-\nu^2}\,. \label{eq33}
\end{equation}
We note that these amplitudes have no kinematic but only dynamic singularities
(poles and cuts). As a consequence, the integrals on the rhs of
Eqs.~(\ref{eq32}) and (\ref{eq33}) yield real and regular functions in a
triangle bounded by the onset of particle production, that is below the cuts
from $\pi\pi$ states at $t\geq 4M_{\pi}^2$ and N$\pi$ states for
$s,u\geq(M_N+M_{\pi})^2$. Therefore, the integrals (``dispersive
contributions'') can be expanded in a Taylor series about the origin of the
Mandelstam plane. In order to avoid the cluttering of indices in the following
text, we denote the real part of the dispersive amplitudes simply by ${\rm
Re}A^I_{i,{\rm disp}}={\rm Re}A^I_{i}-A^I_{i,{\rm pole}}\doteq A_i^I$.
\section{Anomalous magnetic moment}
\begin{figure}[hbt]
\begin{center}
\epsfig{file=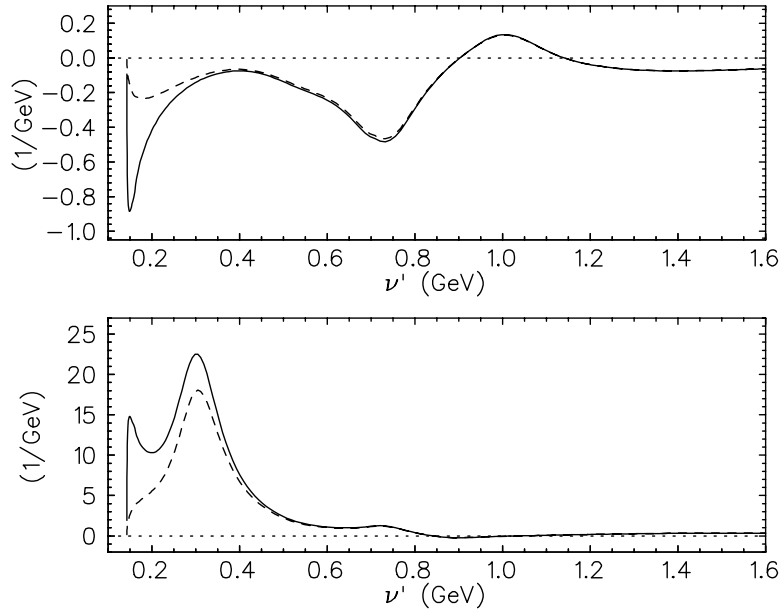,width=8cm,angle=90} \caption[]{The integrands of
the dispersion integrals on the rhs of Eq.~(\ref{eq41}) for the isoscalar (top)
and isovector (bottom) combinations of the amplitudes
$A_1(\nu,t_{\rm{thr}},0)$. The full lines are obtained for
$\nu=\nu_{\rm{thr}}$, the dashed lines for $\nu=0$.} \label{fig41}
\end{center}
\end{figure}

\begin{figure}[htb]
\begin{center}
\epsfig{file=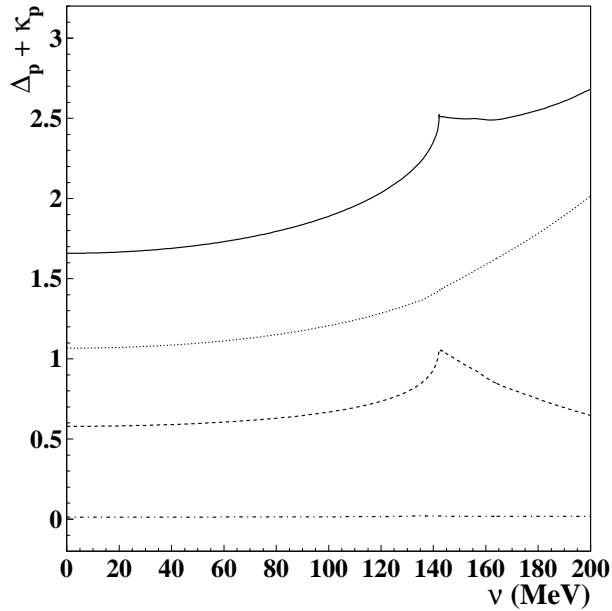,width=8cm, width=8cm} \caption{The values of
$\kappa_p+\Delta_p(\nu,t_{\rm{thr}})$ obtained from the dispersion integral of
Eq.~(\ref{eq41}) with a multipole decomposition of Im$A_1$. Solid line: full
result for Im$A_1$ evaluated with MAID03. Dashed line: result for the
dispersion integral with only the S-wave contribution to Im$A_1$, dotted line:
P-wave contribution only, dashed-dotted line: sum of D- and F-wave
contributions.} \label{fig42}
\end{center}
\end{figure}

The FFR prediction for the Pauli form factor can be cast in the following form:
\begin{equation}
F_2^N(Q^2) \tau_3 + \Delta_N (\nu, t_{\rm{thr}},Q^2) = \frac{4 M_N^2}{\pi e
g_{\pi N}} {\cal P}\int_{\nu_{\rm{thr}}}^{\infty}{\rm d}\nu' \,
\frac{\nu'\,{\rm Im}\,A_1^{(N, \pi^0)}(\nu',t_{\rm{thr}},Q^2)}{\nu'^2-\nu^2}\,,
\label{eq41}
\end{equation}
where $\Delta^N(\nu, t, Q^2)$ is the ``FFR discrepancy'', which should vanish
in the soft-pion limit, $\nu\rightarrow 0, \nu_B\rightarrow 0,
M_{\pi}\rightarrow 0$. In this limit, the pion threshold moves to the point
($\nu=0$, $t=-Q^2$) of the Mandelstam plane, which lies in the unphysical
region. In the following we concentrate on the question whether the anomalous
magnetic moment $\kappa_N=F_2^N(0)$ can be determined from the photoproduction
data. The integrand for the dispersion integral of Eq.~(\ref{eq41}) is plotted
in Fig.~\ref{fig41} for real photons ($Q^2=0$) and at $\nu=0$ (dashed lines) as
well as $\nu=\nu_{\rm{thr}}$ (solid lines). In accordance with the magnetic
moments, the isoscalar combination (top) is small compared to the isovector one
(bottom).
The isoscalar integrand (top) shows peaks at threshold (S-wave pion
production), no contribution in the $\Delta(1232)$ region, and further peaks in
the second and third resonance regions. The isovector integrand (bottom) is
essentially given by S-wave threshold production and a large contribution of
the $\Delta(1232)$, whereas the higher resonance regions are negligible.
However we note that the size of the S-wave contribution decreases strongly if
the integral is evaluated in sub-threshold kinematics, for example, at
$\nu=0$ as shown by the dashed lines in the figure.\\
\indent In Fig.~\ref{fig42} we investigate the convergence of the multipole
expansion for the dispersion integral of the proton. The figure shows the lhs
of Eq.~(\ref{eq41}) evaluated over a large energy range. The result is clearly
dominated by the imaginary part of the P-wave amplitude (dotted line), but the
S-wave contribution is substantial at low $\nu$ values and yields the cusp
effect at threshold (dashed line). The imaginary parts of the higher partial
waves turn out to be negligible over the whole energy region.
We observe that the FFR theorem is nearly fulfilled in the sub-threshold
region, for $\nu \rightarrow 0$. Indeed, the result of the dispersion integral
is very close to $\kappa_p=1.793$, and by increasing $t$ from $t_{\rm{thr}}$ to
$M_{\pi}^2$, the FFR discrepancy decreases even further. In this sense we may
conclude that the proton's anomalous magnetic moment is produced to
$\frac{1}{3}$ by S-wave pions near threshold and  $\frac{2}{3}$ by P waves,
mostly from $\Delta(1232)$ resonance excitation. The following Fig.~\ref{fig43}
compares the predictions of Heavy Baryon (HB) ChPT and DR for $\Delta_p
(\nu,t_{\rm{thr}},0)$ as function of $\nu$ in the range $0\le\nu\le200$~MeV.
Although there is good agreement in the threshold region (upper panel), we
observe 3 principal differences between HBChPT and DR if we move further away
from the threshold (lower panel): (I) The rise of $\Delta_p$ for
$\nu\gtrsim170$~MeV due to the $\Delta(1232)$ resonance cannot be described by
the ``static'' LECs of HBChPT, but requires a dynamical description of the
resonance degrees of freedom ~\cite{Gai05,PV05}. (II) The curvature predicted
by HBChPT for small $\nu$ is due to the non-relativistic approach, which leads
to shifts of the nucleon pole situated at $\nu=\nu_B=-9.7$~MeV and, along the
same line, to violations of the crossing symmetry. (III) Different order
approximations of HBChPT lead to quite different results in the sub-threshold
region.
\begin{figure}[htb]
\begin{center}
\epsfig{file=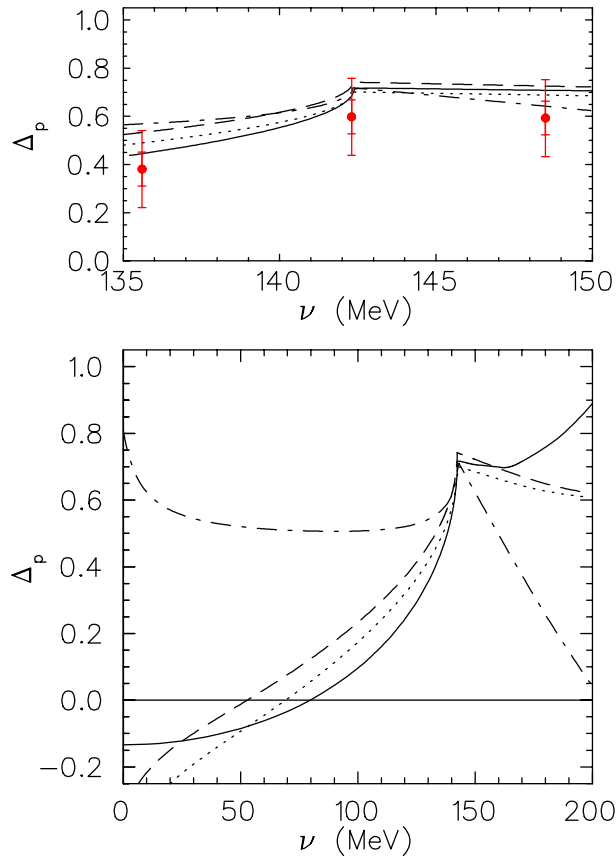, width=8cm } \caption[]{The correction to the FFR
sum rule for the proton, $\Delta_p(\nu,t_{\rm{thr}},0)$ as defined by
Eq.~(\ref{eq41}). The dispersive approach (solid lines) is compared to several
predictions of HBChPT, represented by the dotted~\cite{BKM96},
dashed~\cite{BKM96a}, and dashed-dotted~\cite{Ber01} lines. The data points are
obtained from the experimental S- and P-wave multipoles of Ref.~\cite{Sch01}
and an estimate of the D waves according to MAID.}\label{fig43}
\end{center}
\end{figure}
The shortcomings of HBChPT have of course been noted often before, and several
groups are now applying newly developed manifestly Lorentz-invariant
renormalization schemes~\cite{Bec99,Kub01,Fuc03} to various physical processes,
in particular also to pion photoproduction. However, the general structure of
the dispersive part of the (dispersive) amplitude $A_1$ for small external
momenta has already been given in Ref.~\cite{Ber92a},
\begin{equation}
A_1^{(N,\pi^0)} = a_{00} + a_{02}\nu_B(t) + a_{20}\nu^2 + \ldots\,,
\label{eq42}
\end{equation}
where the coefficients $a_{ik}$ are functions of the mass ratio $\mu$. In
particular the leading coefficients depend on the pion mass as follows:
$a_{00}={\mathcal O}\,(\mu^2)$, $a_{02}={\mathcal O}\,(\ln\,\mu)$, and
$a_{20}={\mathcal O}\,(\mu^{-1})$. The vanishing of $a_{00}$ in the chiral
limit is, of course, a necessary condition for the validity of the FFR sum
rule. Furthermore, the divergence of the higher expansion coefficients in that
limit is the reason why the old LET for neutral pion photoproduction failed.
The analytic continuation of the multipole expansion for $A_1$ to the soft-pion
threshold at the unphysical point ($s=u=M_N^2, t=M_\pi^2$) requires some care,
because the Legendre polynomials $P_\ell(x)$ involved in the expansion have to
be evaluated at $|x|>1$. As a result of this extrapolation we find that
$\delta_{00}^N$ is zero within the error bars of our calculation, which are
essentially due to the unknown higher part of the spectrum. Also the expansion
in $\nu_B(t)$ requires an extrapolation to the unphysical region near the
origin of the Mandelstam plane. However, the coefficients of $\nu^{2n}$,
$n\geq1$, can be (approximately) obtained by expanding the dispersion integral
at $t=t_{\rm {thr}}$ in a power series in $\nu^2$,
\begin{equation}
a_{2n,0}^N \approx \frac{4M_N^2}{\pi e g_{\pi N}}\,\int_{\nu_{\rm{thr}}}^\infty
\frac{d\nu'}{(\nu')^{2n+1}}\,{\rm Im}\,A_1^{(N,\pi^0)}(\nu',t_{\rm{thr}},0) \,.
\label{eq43}
\end{equation}
Due to the additional factors of $1/\nu'\,^2$, these integrals are well
saturated by the region between threshold and the $\Delta(1232)$ resonance. The
numerical results for these coefficients are $a_{20}^p=0.368/M^2_{\pi+}$,
$a_{40}^p=0.120/M^4_{\pi+}$, and $a_{60}^p=0.054/M^6_{\pi+}$, and
Fig.~\ref{fig44} shows the good convergence of the Taylor series below pion
threshold.
\begin{figure}[htb]
\begin{center}
\epsfig{file=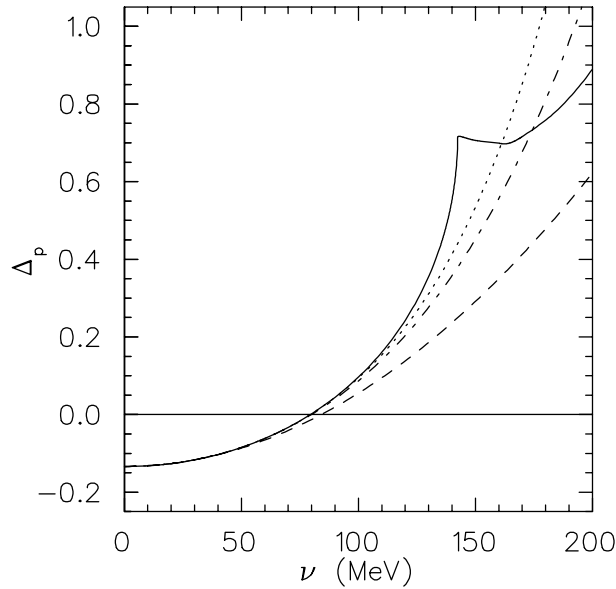, width=8cm } \caption{Full line: the correction
$\Delta_p(\nu,t_{\rm{thr}},0)$ as defined by Eq.~(\ref{eq41}), compared to the
low-energy expansion in the crossing-symmetric variable $\nu$. Dashed line:
expansion up to ${\mathcal O} (\nu^2)$, dashed-dotted line: to ${\mathcal O}
(\nu^4)$, dotted line: to ${\mathcal O} (\nu^6)$.}\label{fig44}
\end{center}
\end{figure}
We note that these results are obtained by expanding the denominator
$\nu'^2-\nu^2$ in the dispersion integral for fixed $t_{\rm{thr}}$ and $Q^2=0$.
This expansion converges up to the cusp at $\nu=\nu_{\rm{thr}}$. In the same
way, also the loop terms of relativistic ChPT can be expanded in a real power
series in $\nu^2$ about $\nu=0$, up to the singularity at threshold. The Born
and counter terms are power series in $\nu^2$ anyway. It is therefore possible
to determine the (unknown) low-energy constants by comparing the
Taylor expansions of ChPT and DR in the sub-threshold region.\\
\begin{figure}[hbt]
\begin{center}
\epsfig{file=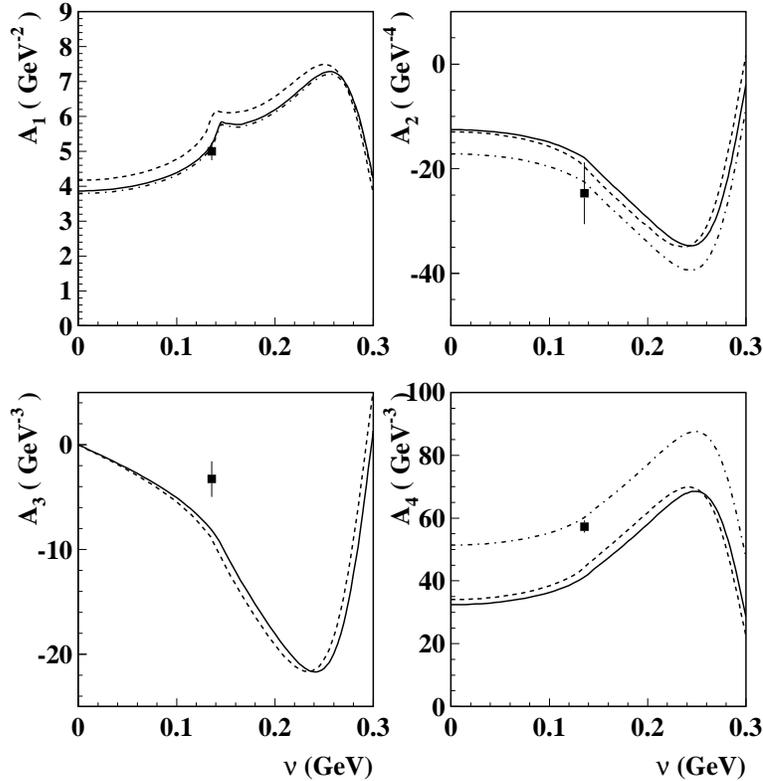,width=10cm} \caption[]{The real parts of the
amplitudes $A_i^{(p\pi^0)}$ for the reaction $\gamma p\to\pi^0 p$ as function
of $\nu$ and at $t=t_{\rm{thr}}$. Solid lines: dispersive contributions
according to Eqs.~(\ref{eq32}) and (\ref{eq33}) as evaluated with the imaginary
amplitudes from MAID05 containing partial waves up to $\ell_{\rm{max}}=3$.
Dashed lines: same results calculated with SAID~\cite{VPI}. The dashed-dotted
lines are obtained by adding the vector-meson contributions to the MAID result.
The data points near threshold are derived from the experimental values of
Ref.~\cite{Sch01} for the S and P waves plus the MAID values for the D waves.}
\label{fig45}
\end{center}
\end{figure}
\indent Figure~\ref{fig45} displays the 4 invariant amplitudes for neutral pion
photoproduction on the proton. The agreement between the MAID and SAID results
demonstrates that the imaginary parts of the amplitudes, which serve as input
for the dispersion integrals, are quite similar in these two partial-wave
analyses. We further observe that the cusp effect is only dominant for the
amplitude $A_1$, which signifies that the other 3 amplitudes have only small
contributions from S waves and loop effects. However, the dispersion integrals
can not fully describe the experimental values for $A_2$ to $A_4$. The reason
is that these integrals do not provide the pole structures of the vector mesons
at $t=m_V^2$, even though the vector meson background plays an important role
in the unitarization process of MAID. It is even more surprising that the
dispersion integrals for the threshold amplitudes change only by a few percent
if we drop the vector mesons in the construction of the absorptive amplitude by
MAID, whereas the vector mesons yield 20~\% and 50~\% of the threshold
amplitudes for $A_2$ and $A_4$, respectively. We therefore have to accept the
Mandelstam hypothesis~\cite{Man58} that the amplitudes are the sum of all pole
terms plus two-dimensional integrals over the double spectral region. The
one-dimensional DRs, e.g. at $t=$const, follow from this representation, as has
been proved for pion photoproduction by Ball~\cite{Bal61}. Alternatively, we
could subtract the DRs at $\nu=0$, which introduces an unknown function
$A_i^I(0,t,0)$. This function is real in the region of small $t$, and in
principle can be constructed from its imaginary part by an integral along the
$t$-axis of Fig.~\ref{fig31}. If we add the t-channel $\rho$ and $\omega$ poles
according to MAID05, we obtain an almost perfect agreement for $A_1$, $A_2$ and
$A_4$. The apparent discrepancy between theory and experiment for $A_3$ is an
open question. The inclusion of the vector meson poles does not help, because
only axial vector mesons can contribute to the crossing-odd amplitude
$A_3^{(p\pi^0)}$~\cite{Pas06}.\\
\indent Let us next expand all 4 amplitudes for neutral pion photoproduction
about the point ($\nu=0,\,t=M_{\pi}^2,Q^2=0$). We first define dimensionless
quantities $\Delta_i(\nu,t,0)$ as follows:
\begin{eqnarray}\label{44}
A_1(\nu,t,0)& = &\frac{eg_{\pi N}}{2M_N^2}\,( \kappa
+\Delta_1(\nu,t,0)) \,, \nonumber\\
A_2(\nu,t,0) & = & \frac{eg_{\pi N}}{2M_N^4}\,\Delta_2(\nu,t,0)\,, \\
A_{3,4}(\nu,t,0) & = & \frac{eg_{\pi N}}{2M_N^3}\,\Delta_{3,4}(\nu,t,0)\,.
\nonumber
\end{eqnarray}
The functions $\Delta_i(\nu,t,0)$ are regular near the origin of the Mandelstam
plane and can be expanded in a power series in $\nu$ and $t$ (or $\nu_B(t)$).
As is evident from the definition of the variables, $\nu$ is
${\mathcal{O}}(M_\pi)$ and $\nu_B$ is ${\mathcal{O}}(M_\pi^2/M_N)$ in the
region of interest. Therefore the crossing-even amplitudes
$\Delta_{1,2,4}^{(+,0)}$ and $\Delta_{3}^{(-)}$ have the expansion
\begin{equation}
\Delta_i(\nu,t,0)  = \delta_{00}^i +\delta_{20}^i\,\nu^2/M_\pi^2 +
\delta_{02}^i\,\nu_B/M_\pi + \ldots \,, \label{45}
\end{equation}
with the lowest expansion parameters given by
\begin{eqnarray}
\delta_{00}^i & = & \Delta_i(0,\,M^2_\pi,0)\,, \nonumber\\
\delta_{20}^i & = &
\frac{M_\pi^2}{2}\frac{\partial^2}{\partial\,\nu^2}\,\Delta_i(\nu\,,M^2_\pi,0)\big|_{\nu=0}\,, \nonumber \\
\delta_{02}^i & = &
4M_N\,M_\pi\frac{\partial}{\partial\,t}\,\Delta_i(0,t,0)\big|_{t=M^2_\pi} \,.
\label{46}
\end{eqnarray}
The expansion of the crossing-odd amplitudes $\Delta_{3}^{(+,0)}$ and
$\Delta_{1,2,4}^{(-)}$ takes the form
\begin{equation}
\Delta_i(\nu,t,0) = \delta_{10}^i\,\nu/M_\pi + \delta_{30}^i\,\nu^3/M_\pi^3 +
\delta_{12}^i\,\nu\nu_B/M_\pi^2 + \ldots \,, \label{eq47}
\end{equation}
with the lowest expansion parameter
\begin{equation}
\delta_{10}^i =
M_\pi\frac{\partial}{\partial\nu}\,\Delta(\nu,M_\pi^2)\big|_{\nu=0}\,.
\label{eq48}
\end{equation}
\begin{table}[htb]
\begin{center}
\caption{The leading expansion coefficients for the $p\pi^0$ amplitudes from
the dispersion integral (see Eqs.~(8)-(\ref{eq48}) for definitions) including
the vector meson $t$-channel contributions. In brackets: the low-energy
constants of ChPT~\cite{Ber05}.} \label{tab1}
\begin{tabular}{|c|cccc|}
\hline
& $\delta_{00}$ & $\delta_{10}$ & $\delta_{20}$ & $\delta_{02}$\\
\hline
$A_1$ & -0.04 \,(0) &  -  &  0.32 \,(0.53) &  1.68 \,(3.40) \\
$A_2$ & -6.41 \,(-6.33) & - & -1.26 & 1.92 \\
$A_3$ & - & -2.23 (-2.58) &  - & -\\
$A_4$ & 21.19  \,(22.40) & - & 2.23 & 4.50 \\
\hline
\end{tabular}
\end{center}
\end{table}
In Table~\ref{tab1} we list the leading expansion coefficients for
the full $p\pi^0$ amplitudes obtained from the dispersive
calculation. The numbers in brackets are the LECs of relativistic
ChPT. The differences between the full dispersive result and the
LECs indicate the size of the loop terms and, in the case of $A_1$,
also of the FFR current. It is obvious that the loop contributions
are small for the amplitudes $A_2$, $A_3$, and $A_4$.
\section[]{Pion electroproduction}
The threshold for pion electroproduction moves with the value of $Q^2$, and in
the following we evaluate the dispersive amplitude $t = t_{\rm{thr}}(Q^2)$
along the path from $\nu = \nu_{\rm{thr}}(Q^2)$ to infinity. In the soft-pion
limit, the threshold moves to $\nu=0$ and $t=-Q^2$ (or $\nu_B = 0$). As in the
previous section for real photons, we can extrapolate from the physical to the
soft-pion threshold for small values of $\nu$, $\nu_B$, and $Q^2$. Of course,
we can not expect to reproduce the FFR sum rule in this way, because the
expansion coefficients of the FFR discrepancy depend on the pion mass and the
dispersion calculation only provides these coefficients for the physical mass.
In particular the pion-loop effects at threshold depend on the pion mass and,
moreover, they produce a $Q^2$ dependence very different from the nucleon form
factors. However, as has been shown before, we expect a suppression of these
loop effects if the dispersion integral is evaluated in the sub-threshold
region.
\vspace{-2cm}
\begin{figure}[htb]
\begin{center}
\epsfig{file=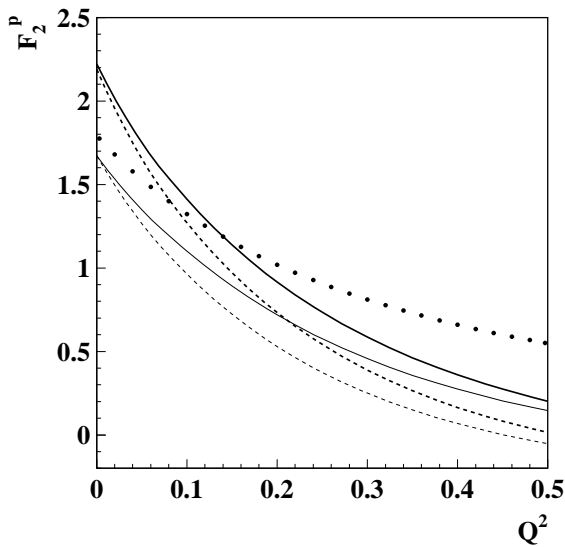,width=10cm}\vspace{-1cm}
\caption[]{The proton's Pauli form factor $F_2^p$ as function of $Q^2$ compared
to the rhs of Eq.~(\ref{eq41}). Thick solid line: dispersive results at
threshold ($\nu=\nu_{\rm{thr}}, t=t_{\rm{thr}}$), thick dashed line: same
kinematics but including the $t$-channel vector meson poles, thin solid line:
dispersive results for soft-pion kinematics ($\nu=\nu_B=0$), thin dashed line:
same kinematics but including the $t$-channel vector meson poles. The
dashed-dotted line is the parametrization of $F_2^p$ according to
Ref.~\cite{Kel04}.}\label{fig51}
\end{center}
\end{figure}
In Fig.~\ref{fig51} we compare the Pauli form factor $F_2^p$ (dotted line) to
the $Q^2$ dependence of $A_{1, \,\rm{disp}}^{(p \pi^0)}$ as evaluated by the
dispersion integral at $\nu=0$ and $\nu = \nu_{\rm{thr}}$. The figure clearly
demonstrates that the slopes of the Pauli form factor and the invariant
amplitude $A_{1,\,\rm{disp}}^{(p \pi^0)}$ differ considerably. More
quantitatively, the extrapolation to the soft-pion kinematics yields an
effective r.m.s. radius $r[A_1(\nu = \nu_B=0)] \approx 1.12$~fm, much larger
than the Pauli radius of the proton, $r_2^p = 0.894$~fm~\cite{Mer96} or
$0.879$~fm~\cite{Kel04}. The reason for this behavior is already seen from the
integrand of the dispersion integral shown in the top panels of
Fig.~\ref{fig52} for the momentum transfers $Q^2=0$ and $0.1$~GeV$^2$.
Evidently the bulk contribution to the integral stems from the $\Delta(1232)$
resonance. In the real photon limit and for energies near threshold (solid
line) also the S-wave threshold production is quite sizeable, but this
contribution of the pion cloud decreases rapidly if the energy moves into the
sub-threshold region. It is also seen that the loop effects drop faster with
momentum transfer than the resonance contributions. As discussed in the
previous section, the FFR prediction is essentially verified at the real photon
point. This fact is also seen from Fig.~\ref{fig52} by comparing the
sub-threshold results (thin lines) with the Pauli form factor (dotted line) at
$Q^2=0$.
\begin{figure}[htb]
\begin{center}
\epsfig{file=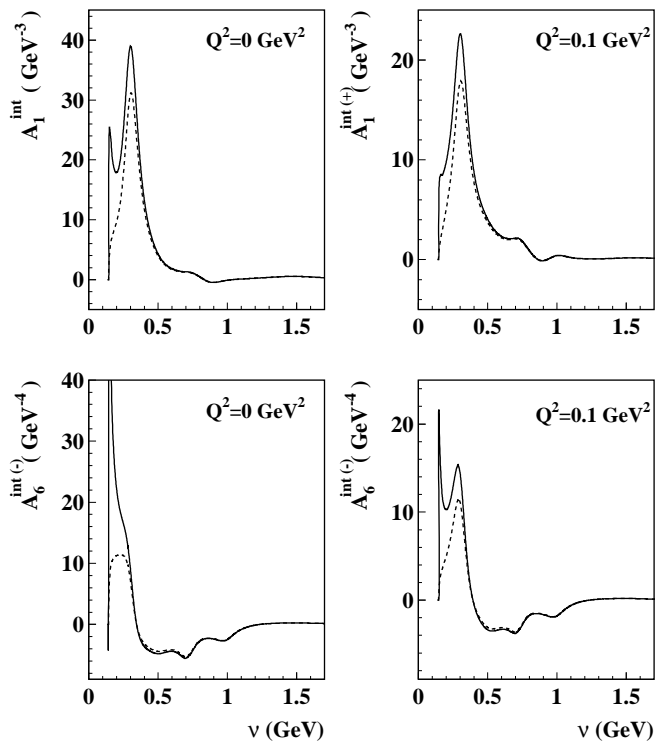,width=10cm} \caption[]{The integrands
of the dispersion integrals for $A_1^{(p \pi^0)}$ (top row) and $A_6^{(-)}$
(bottom row) as function of the integration variable $\nu'$. Left: $Q^2=0$,
right: $Q^2=0.1$~GeV$^2$; solid lines: integrands for the threshold amplitude
($\nu= \nu_{\rm{thr}}$), dashed lines: integrands for the sub-threshold
amplitude ($\nu=0$).}\label{fig52}
\end{center}
\end{figure}
On the other hand, the figure shows that the slopes of the amplitudes differ
considerably from the slope of $F_2^p$ even in the sub-threshold region. In a
simple model including the S-wave loop contributions according to
Ref.~\cite{Ber91} plus the FFR contributions for all the multipoles, we obtain
the following slope for the threshold amplitude and its contributions, all in
units of GeV$^{-4}$:
\begin{eqnarray}
\frac {\partial}{\partial
Q^2}A_{1,\,\rm{disp}}^{(p\pi^0)\,\rm{thr}}(\rm{model})&=&
-311~(E_{0+})+ 278~(L_{0+})-1~({\rm{P~waves}})\nonumber \\
&=&-13.5~({\rm{FFR}})-20.5~({\rm{loop}})=-34~{\rm{GeV}}^{-4} \, . \label{eq51}
\end{eqnarray}
The strong cancelation between the transverse and the longitudinal S-wave
contributions is remarkable, it requires a good knowledge of both multipoles to
get a reliable prediction for the slope. Furthermore, the slope is largely
determined by the loop contribution. Translated into transition radii, the FFR
term has the radius of the Pauli form factor, $r_2^p \approx 0.88$~fm, whereas
the pion cloud reaches to much larger distances described by $r[{\rm
{loop}}]\approx 1/M_{\pi} = 1.45$~fm. The total result is $r[{\rm {model}}] =
1.12$~fm, in good agreement with $r[\nu_{\rm{thr}},\, t_{\rm{thr}}] = 1.17$~fm
and $r[\nu=0,\, t_{\rm{thr}}] = 1.09$~fm obtained from the dispersion
integrals. In conclusion, the radius derived from the invariant amplitude $A_1$
is about 25~\% larger than the Pauli radius, which is another "smoking gun" for
the importance of the pion cloud in low-energy nuclear physics.\\
\indent Let us now turn to the second FFR sum rule, which connects the axial
and Dirac isovector form factors with the amplitude $A_6^{(-)}(\nu,
t_{\rm{thr}}, Q^2)$. Its physics content is identical with the LET of Nambu
{\it et al.}~\cite{Nam62}, which has been derived for the slope of the S-wave
multipole. In the notation of Fubini {\it et al.}~\cite{Fub65} this sum rule
takes the following form in the soft-pion limit:
\begin{equation}
\label{eq52} \frac{2}{\pi} {\cal P}\int_{\nu_{thr}}^{\infty}{\rm d}\nu'\,
\frac{\nu'\,{\rm Im}\,A_6^{(-)}(\nu',t_{thr},Q^2)}{\nu'^2-\nu^2}\rightarrow
\frac{eg_{\pi N}}{2M_NQ^2}\,[G_A^V(Q^2)-F_1^V(Q^2)].
\end{equation}
The isovector Dirac radius is relatively well known from the analysis of
elastic electron scattering, e.g., $\langle r^2\rangle_1^V =
(0.585\pm0.010)$~fm$^2$~\cite{Mer96}. The axial mass parameter as determined by
neutrino and antineutrino scattering~\cite{Ahr88} lies in the range of
$M_A=(1.026\pm0.021)$~GeV corresponding to $\langle r^2\rangle_A^V
=(0.444\pm0.019)$~fm$^2$~\cite{Lie99}, which leads to
$A_{6,\rm{FFR}}^{(-)}(Q^2=0)=(1.30\pm0.27)~\rm{GeV}^{-3}$. Alternatively, the
same axial mass but the radius of Ref.~\cite{Kel04} results in $(1.55 - 9.50 \,
Q^2/~\rm{GeV}^2+{\cal O} (Q^4)) \,~\rm{GeV}^{-3}$. In the soft-pion limit only
the S-wave multipole $E_{0+}(Q^2)$ survives as long as $Q^2$ is finite, and in
accordance with the LET of Nambu {\it et al.}~\cite{Nam62}, the information of
the LET resides in the slope of that multipole. For the real pion mass, on the
other hand, the multipole $L_{0+}^{(-)}$ accounts for 94~\% of
$A_{6,\,\rm{FFR}}^{(-)}$ at $Q^2=0$, the remainder being given by
$L_{1-}^{(-)}$. However, already at $Q^2=0.1$~GeV$^2$ the bulk contribution
(78~\%) is due to the rising  multipole $E_{0+}$. Obviously, the situation is
more complicated in the real world than for massless pions. The integrand for
the amplitude $A_{6,\, \rm{disp}}^{(-)}$ is shown in the lower panels of
Fig.~\ref{fig52}. We note that the integrand diverges at the onset of the
imaginary part like $1/\sqrt{\nu-\nu_{\rm{cusp}}}$. The figure shows positive
contributions from both threshold pion production and $\Delta(1232)$ resonance
excitation. However, these contributions are largely canceled by equally strong
ones with opposite sign in the second and third resonance regions. It is again
seen that the S-wave loop effects drop very much faster with momentum transfer
$Q^2$ than the resonance contributions do. The multipole decomposition of the
threshold amplitude for $Q^2=0$ takes the form
\begin{eqnarray}
A_{6,\rm{thr}}^{(-)}&\approx & 3.8~(E_{0+})
-1.5~(L_{0+})+0.9~(M_{1+})+1.0~(E_{1+})-1.0~(L_{1+})
\nonumber\\
&&-0.4~(L_{1-})-1.7~(E_{2-})+0.4~(L_{2-})- 0.2~({\rm{others}})\nonumber\\
&\approx & 2.3~({\rm{S}})+0.7~({\rm{P}})-1.3~({\rm{D}})-0.4~(\rm{F}) \approx
1.3\, ,\label{eq53}
\end{eqnarray}
all in units of GeV$^{-3}$. We observe a terrific cancelation among
the multipoles of the same pion partial wave and also between the
strong electromagnetic dipole excitations $E_{0+}$ and $E_{2-}$. It
is surprising to see that the electric transverse and longitudinal
multipoles of the $\Delta(1232)$ resonance, $E_{1+}$ and $L_{1+}$,
contribute just as much as the magnetic $M_{1+}$ transition,
although the latter multipole is stronger by a factor of 40 and 25,
respectively. As is also seen in Fig.~\ref{fig52}, both S and P
waves yield positive contributions, whereas the D and F waves of the
second and third resonance regions diminish the integral. As $\nu$
moves from $\nu_{\rm {thr}}$ to zero, the total S-wave contribution
decreases considerably, whereas the higher multipole contributions
change little. As a result the invariant amplitude becomes negative
in the sub-threshold region. We conclude that the dispersive
threshold amplitude is right on top of the sum rule value. However,
MAID is based on an isospin-symmetric form of the production
amplitude, and therefore the S-wave amplitudes near threshold are
only approximately correct. In view of the discussed cancelations,
the perfect agreement is therefore sheer chance.
\section{Summary and Conclusions}
We have studied pion photo- and electroproduction on the basis of dispersion
relations at $t$=const, analyzing the invariant amplitudes in both the
(unphysical) sub-threshold region and the physical region between threshold and
the $\Delta(1232)$ resonance. Our findings may be summarized as follows:
\begin{itemize}
\item{The extension to sub-threshold kinematics provides a unique framework
to determine the low-energy constants of chiral perturbation theory by global
properties of the excitation spectrum.}
\item {The Fubini-Furlan-Rossetti sum rule allows us to determine the anomalous
magnetic moments of the nucleons, $\kappa_p$ and $\kappa_n$, from the pion
photoproduction amplitude $A_1^{(N\pi^0)}$. In particular, $\kappa_p$ is
related to $\Delta(1232)$ excitation ($\frac {2}{3}$) and S-wave pion
production near threshold ($\frac {1}{3}$).}
\item {The predictions of Fubini, Furlan, and Rossetti for the nucleon form factors
are violated by the chiral symmetry breaking due to the finite pion
mass.  The resulting pion-loop corrections have transition radii far
above the nucleon radius. For the same reason the curvature of the
invariant amplitudes is very much larger than predicted by the sum
rule. As a consequence, the terms ${\mathcal {O}}~(Q^4)$ are of
``unnaturally'' large size such that the convergence radius of a
one-loop expansion is expected to be very small.}
\item {The comparison between the SAID and MAID results in Fig.~\ref{fig45}
shows that the input of the dispersion relations, the imaginary parts of the
amplitudes, are more constrained by the experimental data than the real parts
are.}
\end{itemize}
We conclude that dispersion relations allow us to construct a unitary, gauge
and Lorentz invariant description of pion production based on experimental
information for the absorptive part of the amplitudes. The agreement found
between the experimental threshold amplitudes and the dispersive results is
generally satisfactory. We hope to extend these calculations to the higher
energies up to the $\Delta(1232)$ resonance in order to constrain the real part
of the background multipoles, which still hampers the model-independent
determination of the small electric and Coulomb amplitudes for $\Delta(1232)$
excitation.
\section*{Acknowledgements}
This work was supported by the Deutsche Forschungsgemeinschaft (SFB
443) and the EU ``Integrated Infrastructure Initiative Hadron
Physics'' Project under contract number RII3-CT-2004-506078. We are
grateful to DFG and ROC for supporting our common research through
the SFB~443, by joint project NSC/DFG 446 TAI113/10/0-3 in the field
of hadron physics over the past decade. Our particular thanks go to
Shin-Nan Yang for a fruitful collaboration and a generous
hospitality.

\end{document}